\documentclass{svjour3}                     
\smartqed  
\usepackage{graphicx}
\usepackage{amssymb}
\usepackage[mathscr]{eucal}
\usepackage{graphicx}

\usepackage{layout}
\usepackage{lipsum}
\usepackage{enumerate}
\usepackage{amssymb}

\usepackage{mathptmx} 
\usepackage{latexsym}
\usepackage{booktabs}
\usepackage{multirow}
\usepackage{tabularx}
\usepackage{caption}
\usepackage[figuresright]{rotating}
\usepackage{latexsym}
\usepackage{graphicx}
\usepackage[mathscr]{eucal}
\usepackage{amsmath}
\usepackage{mathtools}
\usepackage{booktabs}

\usepackage{array,booktabs,tabularx}

\usepackage{amsmath}
\usepackage{mathtools}

\usepackage[rightcaption]{sidecap}

\usepackage{graphicx} 
\graphicspath{ {images/} }

\usepackage{mathptmx}      
\usepackage{latexsym}
\usepackage{booktabs}
\usepackage{multirow}
\usepackage{tabularx}
\usepackage{caption}
\usepackage{latexsym}

 \usepackage{mathptmx}      
\usepackage{latexsym}
\usepackage{booktabs}
\usepackage{multirow}
\usepackage{tabularx}
\usepackage{caption}
\begin{document}

\title{The  Laplace-Adomian Decomposition Method Applied to the Kundu-Eckhaus Equation }


\author{O. Gonz\'alez-Gaxiola }


\institute{O. Gonz\'alez-Gaxiola \at
	Departamento de Matem\'aticas Aplicadas y Sistemas, Universidad Aut\'onoma Metropolitana-Cuajimalpa. Vasco de Quiroga 4871, Santa Fe, Cuajimalpa, 05300, Mexico D.F., Mexico\\
	\email{ogonzalez@correo.cua.uam.mx}  }


\maketitle

\begin{abstract}
The  Kundu-Eckhaus equation is a nonlinear partial differential equation which seems in the quantum field theory, weakly nonlinear dispersive water waves and nonlinear optics. In spite of its importance, exact solution to this nonlinear equation are rarely found in literature. In this work, we solve this equation and present a new approach to obtain the solution  by means of the combined use of the  Adomian Decomposition Method and the Laplace Transform (LADM). Besides, we compare the behaviour of the solutions obtained with the only exact  solutions  given in the literature through fractional calculus. Moreover, it is shown that the proposed method is direct, effective and can be used for many other nonlinear evolution equations in mathematical physics.

\end{abstract}

\section{ Introduction}
\noindent Most of the phenomena that arise in the real world can be  described by means of nonlinear  partial and ordinary differential equations  and, in some cases,  by integral or integro-differential equations. However, most of the mathematical methods developed so far,  are only capable to  solve linear differential equations.  In the 1980's, George Adomian (1923-1996) introduced a powerful method to solve nonlinear differential equations. Since then, this method is known as the Adomian decomposition method (ADM) \cite{ADM-0,ADM-1}. The technique is based on a decomposition of a solution of a nonlinear differential equation in a series of functions. Each term of the series is obtained from a polynomial generated by a power series expansion of an analytic function. The Adomian method is very simple in an abstract formulation but the difficulty arises in calculating the polynomials that becomes a non-trivial task. This method has widely been used to solve equations that come from nonlinear models as well as to solve fractional differential equations \cite{Das-1,Das-2,Das-3}. \\
\noindent The Kundu-Eckhaus equation has been studied by many researchers and those studies done through varied and different methods have yielded much information related to the behavior of their solutions. The mathematical structure of Kundu-Eckhaus equation was studied for the first time in \cite{Eck} and \cite{Kundu}. For example, in \cite{Wang} the authors applied B\"{a}cklund transformation for obtaining bright and dark soliton solutions to the Eckhaus-Kundu equation with the cubic-quintic nonlinearity. After much research have been related to the equation by various methods, many of them can be found in \cite{Bek} and  \cite{Tag} and some applications of the equation nonlinear optics can be found in \cite{Kod} and \cite{Cla}. Recently, in \cite{Haci} the authors obtain obtain some new complex analytical solutions to the Kundu-Eckhaus equation which seems in the quantum field theory, weakly nonlinear dispersive water waves and nonlinear optics using improved Bernoulli sub-equation function method.\\
\noindent In the presente work we will utilize the Adomian  decomposition method in combination with the Laplace transform (LADM) \cite{Waz-Lap} to solve the Kundu-Eckhaus equation.  This equation is a nonlinear partial differential equation that, in  nonlinear optics, is used to model some dispersion phenomena. We will decomposed the nonlinear terms of this equation using the Adomian polynomials and then, in combination with the use of the Laplace transform, we will obtain an algorithm to solve the problem subject to initial conditions. Finally, we will illustrate our procedure and the quality of the obtained algorithm by means of the solution of an example in which the Kundu-Eckhaus equation is solved for some initial condition and we will compare the results with previous results reported in the literature.\\
\noindent Our work is divided in several sections. In ``The Adomian Decomposition Method Combined With Laplace Transform'' section, we present, in a brief and self-contained manner, the LADM. Several references are given to delve deeper into the subject and to study its mathematical foundation that is beyond the scope of the present work. In ``The nonlinear Kundu-Eckhaus Equation'' section, we give a brief introduction to the model described by the  Kundu-Eckhaus equation and we will establish that LADM can be use to solve this equation in its nonlinear version. In ``The General Solution of the Nonlinear Kundu-Eckhaus  Equation Through LADM'' and the  ``Numerical Example'', we will show by means of numerical examples, the quality and precision of our method, comparing the obtained results with the only exact solutions available in the literature \cite{Arzu}. Finally, in the ``Conclusion"  section, we summarise  our findings and present our final conclusions.\\
\section{The Adomian Decomposition Method Combined With Laplace Transform}\label{sec-2}
\noindent  The ADM is a method  to solve ordinary and partial nonlinear  differential equations. Using this method is possible to express analytic solutions in terms of a series \cite{ADM-1,Waz-El}. 
In a nutshell,  the method identifies and separates the linear and nonlinear parts of a differential equation. Inverting and applying the highest order differential operator that is contained in the linear part of the equation, it  is possible to express the solution in terms of the rest of the equation affected by the inverse operator.  At this point, the solution is proposed by means of  a series 
with terms that will be  determined and that give rise to the Adomian Polynomials \cite{Waz-0}. The nonlinear part can also be expressed in terms of these polynomials. 
The initial (or the border conditions) and the terms that contain the independent variables will be considered as the initial  approximation. In this way and by means of a  recurrence relations, it is possible to find the  terms of the series that give the approximate solution of the differential equation.\\
\noindent Given a partial (or ordinary) differential equation 
\begin{equation}
Fu(x,t)=g(x,t)\label{eq:y1}
\end{equation}
with initial condition 
\begin{equation}
u(x,0)=f(x)\label{eq:y2}
\end{equation}
where $F$ is a differential operator  that could, in general, be nonlinear and therefore includes some linear and nonlinear terms.\\  
In general, equation (\ref{eq:y1}) could be  written as 
\begin{equation}
L_{t}u(x,t)+Ru(x,t)+Nu(x,t)=g(x,t)\label{eq:y3}
\end{equation}
where $L_{t}=\frac{\partial}{\partial t}$, $R$ is a linear operator that includes  partial derivatives with respect to $x$,   $N$ is a nonlinear operator and  $g$ is a non-homogeneous term that is $u$-independent.\\
Solving for $L_{t}u(x,t)$, we have 
\begin{equation}
L_{t}u(x,t)= g(x,t)-Ru(x,t)-Nu(x,t)\label{eq:y4}.
\end{equation}
\noindent The Laplace transform  $\mathcal{L}$ is an integral transform discovered by Pierre-Simon Laplace and  is a powerful and very useful technique for solving ordinary and partial differential Equations, which transforms the original differential equation into an elementary algebraic equation \cite{Lap}. Before using the Laplace transform combined with  Adomian decomposition method we review some basic definitions and results on it.\\
{\bf Definition 1} Given a function $f(t)$ defined for all $t \geq 0$, the Laplace transform of
$f$ is the function $F$ defined by
\begin{equation}
F(s)=\mathcal{L}\{f(t)\}=\int_{0}^{\infty}f(t)e^{-st}dt
\end{equation}
for all values of $s$ for which the improper integral converges. In particular $\mathcal{L}\{t^n\}=\frac{n!}{s^{n+1}}$ . \\
\noindent It is well known that there exists a bijection between the set of functions satisfying
some hypotheses and the set of their Laplace transforms. Therefore, it is quite natural
to define the inverse Laplace transform of $F(s)$.\\
{\bf Definition 2} Given a continuous function $f(t)$, if $F(s) = \mathcal{L}\{f(t)\}$, then $f(t)$ is
called the inverse Laplace transform of $F(s)$ and denoted $f(t) = \mathcal{ L}^{-1}\{F(s)\}$.\\
\noindent The Laplace transform has the derivative properties:\\
\begin{equation}
\mathcal{L}\{f^{(n)}(t)\}=s^{n}\mathcal{L}\{f(t)\}-\sum_{k=0}^{n-1}s^{n-1-k}f^{(k)}(0),
\end{equation}
\begin{equation}
\mathcal{L}\{t^{n}f(t)\}=(-1)^{n}F^{(n)}(s),
\end{equation}
where the superscript $(n)$ denotes the $n-th$ derivative with respect to $t$ for $f ^{(n)}(t)$, and
with respect to $s$  for $F^{(n)}(s)$.\\
The LADM consists of applying Laplace transform \cite{Waz-Lap} first on both sides of Eq. (\ref{eq:y4}), obtaining
\begin{equation}
\mathcal{L}\{L_{t}u(x,t)\}= \mathcal{L}\{g(x,t)-Ru(x,t)-Nu(x,t).\}\label{eq:y5}
\end{equation}
An equivalent expression to  (\ref{eq:y5}) is
\begin{equation}
su(x,s)-u(x,0)= \mathcal{L}\{g(x,t)-Ru(x,t)-Nu(x,t)\}\label{eq:y6}
\end{equation}
In the homogeneous case, $g(x,t)=0$, we have
\begin{equation}
u(x,s)=\frac{f(x)}{s}-\frac{1}{s}\mathcal{L}\{Ru(x,t)+Nu(x,t)\}\label{eq:y7}
\end{equation}
now, applying the inverse Laplace transform to equation (\ref{eq:y7})
\begin{equation}
u(x,t)=f(x)-\mathcal{L}^{-1}\big[\frac{1}{s}\mathcal{L}\{Ru(x,t)+Nu(x,t)\}\big]. \label{eq:y8}
\end{equation}
\noindent The ADM method proposes a series solution   $u(x,t)$ given by,
\begin{equation}
u(x,t)= \sum_{n=0}^{\infty}u_{n}(x,t)\label{eq:y7-1}
\end{equation}
The nonlinear term $Nu(x,t)$ is given by
\begin{equation}
Nu(x,t)= \sum_{n=0}^{\infty}A_{n}(u_{0},u_{1},\ldots, u_{n})\label{eq:y8-1}
\end{equation}
where   $\{A_{n}\}_{n=0}^{\infty}$ is the so-called Adomian polynomials sequence established in \cite{Waz-0} and \cite{Ba} and, in general, give us term to term:\\
$A_{0}=N(u_0)$\\
$A_{1}=u_{1}N'(u_0)$\\
$A_{2}=u_{2}N'(u_{0})+\frac{1}{2}u_{1}^{2}N''(u_0)$\\
$A_{3}=u_{3}N'(u_{0})+u_{1}u_{2}N''(u_0)+\frac{1}{3!}u_{1}^{3}N^{(3)}(u_{0})$\\
$A_{4}=u_{4}N'(u_{0})+(\frac{1}{2}u_{2}^{2}+u_{1}u_{3})N''(u_0)+\frac{1}{2!}u_{1}^{2}u_{2}N^{(3)}(u_{0})+\frac{1}{4!}u_{1}^{4}N^{(4)}(u_0)$\\
$ \vdots$\\
\noindent Other polynomials can be generated in a similar way. For more details, see \cite{Waz-0} and \cite{Ba} and references therein. Some other approaches  to obtain Adomian's polynomials can be found in  \cite{Duan,Duan1}.\\
\noindent Using  (\ref{eq:y7-1}) and (\ref{eq:y8-1}) into equation (\ref{eq:y8}), we obtain,
\begin{equation}
\sum_{n=0}^{\infty}u_{n}(x,t)= f(x)-\mathcal{L}^{-1}\Big[\frac{1}{s}\mathcal{L}\{R\sum_{n=0}^{\infty}u_{n}(x,t)+\sum_{n=0}^{\infty}A_{n}(u_{0},u_{1},\ldots, u_{n})\}\Big],\label{eq:y10}
\end{equation}
From the equation (\ref{eq:y10}) we deduce the following recurrence formulas
\begin{equation}
\left\{
\begin{array}{ll}
u_{0}(x,t)=f(x),\\
u_{n+1}(x,t)=-\mathcal{L}^{-1}\Big[\frac{1}{s}\mathcal{L}\{Ru_{n}(x,t)+A_{n}(u_{0},u_{1},\ldots, u_{n})\}\Big],\;\; n=0,1,2,\ldots
\end{array}
\right.\label{eq:y11}
\end{equation}
Using  (\ref{eq:y11}) we can obtain an approximate solution of (\ref{eq:y1}), (\ref{eq:y2}) using  
\begin{equation}
u(x,t)\approx \sum_{n=0}^{k}u_{n}(x,t),\;\; \mbox{where} \;\; \lim_{k\to\infty}\sum_{n=0}^{k}u_{n}(x,t)=u(x,t).\label{eq:y12}
\end{equation}
It becomes clear that, the Adomian decomposition method, combined with the Laplace transform needs less work in comparison with  the traditional  Adomian decomposition method.  This method decreases considerably the volume of calculations. The decomposition procedure of Adomian will be  easily set, without linearising the problem. In this approach, the solution is found in the form of a convergent series with easily computed components; in many cases, the convergence of this series is very fast and only a few terms are needed in order to have an idea of how the solutions behave. Convergence conditions of this series are examined by several authors,  mainly in \cite{Y1,Y2,Y3,Y4}. Additional references related to the use of the Adomian Decomposition Method, combined with the Laplace transform, can be found in \cite{Waz-Lap,ADM-2,Khu}.

\section{The  Nonlinear Kundu-Eckhaus Equation}
\label{Kom-1}

\noindent In mathematical physics, the Kundu-Eckhaus equation is a nonlinear partial differential equations within the nonlinear Schr\"{o}dinger class \cite{Eck,Kundu}:
\begin{equation}\label{Kom-x1}
iu_{t}+u_{xx}+2(\vert u\vert ^{2})_{x}u+\vert u\vert ^{4}u=0.
\end{equation}
In the equation (\ref{Kom-x1}) the dependent variable $u(x,t)$ is a complex-valued function of two real variables $x$ and $t$. The equation (\ref{Kom-x1})  is a basic model that describes optical soliton propagation in Kerr media \cite{Por} . The complete integrability and multi-soliton solutions, breather solutions, and various
types of rogue wave solutions associated with the Kundu-Eckhaus equation have been widely reported by many authors \cite{Ank,Ban,Per,Por}. Nevertheless, in optic fiber communications systems, one
always has to increase the intensity of the incident light field
to produce ultrashort (femtosecond) optical pulses \cite{Zha}. In
this case, the simple NLS equation is inadequate to accurately
describe the phenomena, and higher-order nonlinear terms,
such as third-order dispersion, self-steepening, and self-frequency
shift, must be taken into account \cite{Waz-x,Wang, Wang-1}.\\
\noindent Explicitly calculating the derivatives that appear in equation (\ref{Kom-x1}), we obtain
\begin{equation}\label{Kom-x2}
u_{t}=iu_{xx}+2i(\vert u\vert ^{2})_{x}u+i\vert u\vert ^{4}u.
\end{equation}
To make the description of the the problem complete, we will consider some initial condition
$$u(x,0)=f(x)$$
In the following section we will develop an algorithm using  the method described in  section \ref{sec-2} in order to solve  the nonlinear Kundu-Eckhaus  equation (\ref{Kom-x2}) without resort to any truncation or  linearization.
\section{\bf  The General Solution of the Nonlinear Kundu-Eckhaus  Equation Through LADM}
\label{NLC-ADM}
\noindent Comparing (\ref{Kom-x2}) with equation  (\ref{eq:y4}) we have that $g(x,t)=0$, $L_t$ and $R$ becomes:
\begin{equation}
L_{t}u=\frac{\partial}{\partial t}u,\;\; Ru= i\frac{\partial^{2}u}{\partial x^2},\;\; \label{Oper-1}
\end{equation}
while the nonlinear term is given by
\begin{equation}\label{Oper-2}
Nu=i[2(\vert u\vert ^{2})_{x}+\vert u\vert ^{4}]u.
\end{equation}
\noindent By using now equation (\ref{eq:y11}) through the  LADM  method we obtain recursively
\begin{equation}
\left\{
\begin{array}{ll}
u_{0}(x,t)=f(x),\\
u_{n+1}(x,t)=\mathcal{L}^{-1}\Big[\frac{1}{s}\mathcal{L}\{Ru_{n}(x,t)+A_{n}(u_{0},u_{1},\ldots, u_{n})\}\Big],\;\; n=0,1,2,\ldots
\end{array}
\right.\label{eq:ADM1}
\end{equation}
Note that, the nonlinear term  
$Nu=i[2(\vert u\vert ^{2})_{x}+\vert u\vert ^{4}]u$ can  be split into three terms  to facilitate calculations
$$N_{1}u=iu^{3}\bar{u}^{2},\quad N_{2}u=2iu^{2}\bar{u}_{x},\quad N_{3}u=2iu_{x}u\bar{u}$$
from this, we will consider the decomposition of the nonlinear terms into Adomian polynomials as
\begin{equation}
N_{1}u=iu^{3}\bar{u}^{2}=\sum_{n=0}^{\infty}P_{n}(u_{0},u_{1},\ldots, u_{n})
\label{eq:ADM2}
\end{equation} 
\begin{equation}
N_{2}u=2iu^{2}\bar{u}_{x}=\sum_{n=0}^{\infty}Q_{n}(u_{0},u_{1},\ldots, u_{n}).
\label{eq:ADM2-1}
\end{equation}  
\begin{equation}
N_{3}u=2iu_{x}u\bar{u}=\sum_{n=0}^{\infty}R_{n}(u_{0},u_{1},\ldots, u_{n}).
\label{eq:ADM2-h}
\end{equation}  
Calculating, we obtain
\begin{align*}
P_{0}&=i\bar{u}_{0}^{2}u_{0}^{3},\\
P_{1}&=3i\bar{u}_{0}^{2}u_{0}^{2}u_{1}+2i\bar{u}_{0}u_{1}u_{0}^{3},\\
P_{2}&=3i\bar{u}_{0}^{2}u_{0}^{2}u_{2}+3i\bar{u}_{0}^{2}u_{0}u_{1}^{2}+6i\bar{u}_{0}\bar{u}_{1}u_{0}^{2}u_{1}+2i\bar{u}_{0}\bar{u}_{2}u_{0}^{3}+i\bar{u}_{1}^{2}u_{0}^{3},\\
P_{3}&=3i\bar{u}_{0}^{2}u_{0}^{2}u_{3}+6i\bar{u}_{0}^{2}u_{0}u_{1}u_{2}+i\bar{u}_{0}^{2}{u}_{1}^{3}+6i\bar{u}_{0}u_{1}u_{0}^{2}u_{2}+6i\bar{u}_{0}\bar{u}_{1}u_{0}u_{1}^{2}+2i\bar{u}_{0}\bar{u}_{3}u_{0}^{3}\\&+3i\bar{u}_{1}^{2}u_{0}^{2}u_{1}+2i\bar{u}_{1}\bar{u}_{2}u_{0}^{3}+6i\bar{u}_{0}\bar{u}_{2}u_{0}^{2}u_{1},\\
P_{4}&=3i\bar{u}_{0}^{2}u_{0}^{2}u_{4}+3i\bar{u}_{0}^{2}u_{1}^{2}u_{2}+6i\bar{u}_{0}^{2}u_{0}u_{1}u_{3}+3i\bar{u}_{0}^{2}u_{0}u_{2}^{2}+6i\bar{u}_{0}\bar{u}_{1}u_{0}^{2}u_{3}+2i\bar{u}_{0}\bar{u}_{1}u_{1}^{3}\\&+12i\bar{u}_{0}\bar{u}_{1}u_{0}u_{1}u_{2}+6i\bar{u}_{0}\bar{u}_{2}u_{0}u_{1}^{2}+6i\bar{u}_{0}\bar{u}_{2}u_{0}^{2}u_{2}+6i\bar{u}_{0}\bar{u}_{3}u_{0}^{2}u_{1}+2i\bar{u}_{0}\bar{u}_{4}u_{0}^{3}\\&+3i\bar{u}_{1}^{2}u_{0}^{2}u_{2}+3i\bar{u}_{1}^{2}u_{0}u_{1}^{2}+6i\bar{u}_{1}u_{2}u_{0}^{2}u_{1}+2i\bar{u}_{1}\bar{u}_{3}u_{0}^{3}+i\bar{u}_{2}^{2}u_{0}^{3},
\end{align*}
$$\vdots$$
\begin{align*}
Q_{0}&=2iu_{0}^{2}\bar{u}_{0x},\\
Q_{1}&= 2iu_{0}^{2}\bar{u}_{1x}+4iu_{0}u_{1}\bar{u}_{0x},\\
Q_{2}&=2iu_{1}^{2}\bar{u}_{0x}+2iu_{0}^{2}\bar{u}_{2x}+4iu_{0}u_{1}\bar{u}_{1x}+4iu_{0}u_{2}\bar{u}_{0x}, \\
Q_{3}&=2iu_{1}^{2}\bar{u}_{1x}+2iu_{0}^{2}\bar{u}_{3x}+4iu_{0}u_{1}\bar{u}_{2x}+4iu_{0}u_{2}\bar{u}_{1x}+4iu_{0}u_{3}\bar{u}_{0x}+4iu_{1}u_{2}\bar{u}_{0x}, \\
Q_{4}&=2iu_{2}^{2}\bar{u}_{0x}+2iu_{1}^{2}\bar{u}_{2x}+4iu_{0}u_{1}\bar{u}_{3x}+4iu_{0}u_{2}\bar{u}_{2x}+4iu_{0}u_{3}\bar{u}_{1x}+4iu_{0}u_{4}\bar{u}_{0x}\\&+4iu_{1}u_{2}\bar{u}_{1x}+4iu_{1}u_{3}\bar{u}_{0x},
\end{align*}
$$ \vdots$$
\begin{align*}
R_{0}&=2iu_{0}\bar{u}_{0}u_{0x},\\
R_{1}&= 2iu_{0}\bar{u}_{0}u_{1x}+2iu_{0}\bar{u}_{1}u_{0x}+2iu_{1}\bar{u}_{0}u_{0x},\\
R_{2}&=2iu_{0}\bar{u}_{0}u_{2x}+2iu_{0}\bar{u}_{1}u_{1x}+2iu_{0}\bar{u}_{2}u_{0x}+2iu_{1}\bar{u}_{0}u_{1x}+2iu_{1}\bar{u}_{1}u_{0x}+2iu_{2}\bar{u}_{0}u_{0x},\\
R_{3}&=2iu_{0}\bar{u}_{0}u_{3x}+2iu_{0}\bar{u}_{1}u_{2x}+2iu_{0}\bar{u}_{2}u_{1x}+2iu_{0}\bar{u}_{3}u_{0x}+2iu_{1}\bar{u}_{0}u_{2x}+2iu_{1}\bar{u}_{1}u_{1x}\\&+2iu_{1}\bar{u}_{2}u_{0x}+2iu_{2}\bar{u}_{0}u_{1x}+2iu_{2}\bar{u}_{1}u_{0x}+2iu_{3}\bar{u}_{0}u_{0x},\\
R_{4}&=2iu_{0}\bar{u}_{0}u_{4x}+2iu_{0}\bar{u}_{1}u_{3x}+2iu_{0}\bar{u}_{2}u_{2x}+2iu_{0}\bar{u}_{3}u_{1x}+2iu_{0}\bar{u}_{4}u_{0x}+2iu_{1}\bar{u}_{0}u_{3x}\\&+2iu_{1}\bar{u}_{1}u_{2x}+2iu_{1}\bar{u}_{2}u_{1x}+2iu_{1}\bar{u}_{3}u_{0x}+2iu_{2}\bar{u}_{0}u_{2x}+2iu_{2}\bar{u}_{1}u_{1x}+2iu_{2}\bar{u}_{2}u_{0x}\\&+2iu_{3}\bar{u}_{0}u_{1x}+2iu_{3}\bar{u}_{1}u_{0x}+2iu_{4}\bar{u}_{0}u_{0x},
\end{align*}
$$ \vdots $$
Now, considering (\ref{eq:ADM2}),  (\ref{eq:ADM2-1}) and (\ref{eq:ADM2-h}), we have
\begin{equation}\label{ADM-Y}
N(u)=\sum_{n=0}^{\infty}A_{n}(u_{0},u_{1},\ldots, u_{n})=\sum_{n=0}^{\infty}\big((P_{n}+Q_{n}+R_{n})(u_{0},u_{1},\ldots, u_{n})\big),
\end{equation}
then, the Adomian polynomials corresponding to the nonlinear part $Nu=i[2(\vert u\vert ^{2})_{x}+\vert u\vert ^{4}]u$  are\\
\begin{align*}
A_{0}&=i\bar{u}_{0}^{2}u_{0}^{3}+2iu_{0}^{2}\bar{u}_{0x}+2iu_{0}\bar{u}_{0}u_{0x}, \\
A_{1}&=3i\bar{u}_{0}^{2}u_{0}^{2}u_{1}+2i\bar{u}_{0}u_{1}u_{0}^{3}+2iu_{0}^{2}\bar{u}_{1x}+4iu_{0}u_{1}\bar{u}_{0x}+2iu_{0}\bar{u}_{0}u_{1x}+2iu_{0}\bar{u}_{1}u_{0x}\\&+2iu_{1}\bar{u}_{0}u_{0x},\\
A_{2}&=3i\bar{u}_{0}^{2}u_{0}^{2}u_{2}+3i\bar{u}_{0}^{2}u_{0}u_{1}^{2}+6i\bar{u}_{0}\bar{u}_{1}u_{0}^{2}u_{1}+2i\bar{u}_{0}\bar{u}_{2}u_{0}^{3}+i\bar{u}_{1}^{2}u_{0}^{3}+2iu_{1}^{2}\bar{u}_{0x}+2iu_{0}^{2}\bar{u}_{2x}\\&+4iu_{0}u_{1}\bar{u}_{1x}+4iu_{0}u_{2}\bar{u}_{0x}+2iu_{0}\bar{u}_{0}u_{2x}+2iu_{0}\bar{u}_{1}u_{1x}+2iu_{0}\bar{u}_{2}u_{0x}+2iu_{1}\bar{u}_{0}u_{1x}\\&+2iu_{1}\bar{u}_{1}u_{0x}+2iu_{2}\bar{u}_{0}u_{0x}, 
\end{align*}
\begin{align*}
A_{3}&=3i\bar{u}_{0}^{2}u_{0}^{2}u_{3}+6i\bar{u}_{0}^{2}u_{0}u_{1}u_{2}+i\bar{u}_{0}^{2}{u}_{1}^{3}+6i\bar{u}_{0}u_{1}u_{0}^{2}u_{2}+6i\bar{u}_{0}\bar{u}_{1}u_{0}u_{1}^{2}+2i\bar{u}_{0}\bar{u}_{3}u_{0}^{3}\\&+3i\bar{u}_{1}^{2}u_{0}^{2}u_{1}+2i\bar{u}_{1}\bar{u}_{2}u_{0}^{3}+6i\bar{u}_{0}\bar{u}_{2}u_{0}^{2}u_{1}+2iu_{1}^{2}\bar{u}_{1x}+2iu_{0}^{2}\bar{u}_{3x}+4iu_{0}u_{1}\bar{u}_{2x}\\&+4iu_{0}u_{2}\bar{u}_{1x}+4iu_{0}u_{3}\bar{u}_{0x}+4iu_{1}u_{2}\bar{u}_{0x}+2iu_{0}\bar{u}_{0}u_{3x}+2iu_{0}\bar{u}_{1}u_{2x}+2iu_{0}\bar{u}_{2}u_{1x}\\&+2iu_{0}\bar{u}_{3}u_{0x}+2iu_{1}\bar{u}_{0}u_{2x}+2iu_{1}\bar{u}_{1}u_{1x}+2iu_{1}\bar{u}_{2}u_{0x}+2iu_{2}\bar{u}_{0}u_{1x}+2iu_{2}\bar{u}_{1}u_{0x}\\&+2iu_{3}\bar{u}_{0}u_{0x},
\end{align*}
\begin{align*}
A_{4}&=3i\bar{u}_{0}^{2}u_{0}^{2}u_{4}+3i\bar{u}_{0}^{2}u_{1}^{2}u_{2}+6i\bar{u}_{0}^{2}u_{0}u_{1}u_{3}+3i\bar{u}_{0}^{2}u_{0}u_{2}^{2}+6i\bar{u}_{0}\bar{u}_{1}u_{0}^{2}u_{3}+2i\bar{u}_{0}\bar{u}_{1}u_{1}^{3}\\&+12i\bar{u}_{0}\bar{u}_{1}u_{0}u_{1}u_{2}+6i\bar{u}_{0}\bar{u}_{2}u_{0}u_{1}^{2}+6i\bar{u}_{0}\bar{u}_{2}u_{0}^{2}u_{2}+6i\bar{u}_{0}\bar{u}_{3}u_{0}^{2}u_{1}+2i\bar{u}_{0}\bar{u}_{4}u_{0}^{3}\\&+3i\bar{u}_{1}^{2}u_{0}^{2}u_{2}+3i\bar{u}_{1}^{2}u_{0}u_{1}^{2}+6i\bar{u}_{1}u_{2}u_{0}^{2}u_{1}+2i\bar{u}_{1}\bar{u}_{3}u_{0}^{3}+i\bar{u}_{2}^{2}u_{0}^{3}+2iu_{2}^{2}\bar{u}_{0x}+2iu_{1}^{2}\bar{u}_{2x}\\&+4iu_{0}u_{1}\bar{u}_{3x}+4iu_{0}u_{2}\bar{u}_{2x}+4iu_{0}u_{3}\bar{u}_{1x}+4iu_{0}u_{4}\bar{u}_{0x}+4iu_{1}u_{2}\bar{u}_{1x}+4iu_{1}u_{3}\bar{u}_{0x}\\&+2iu_{0}\bar{u}_{0}u_{4x}+2iu_{0}\bar{u}_{1}u_{3x}+2iu_{0}\bar{u}_{2}u_{2x}+2iu_{0}\bar{u}_{3}u_{1x}+2iu_{0}\bar{u}_{4}u_{0x}+2iu_{1}\bar{u}_{0}u_{3x}\\&+2iu_{1}\bar{u}_{1}u_{2x}+2iu_{1}\bar{u}_{2}u_{1x}+2iu_{1}\bar{u}_{3}u_{0x}+2iu_{2}\bar{u}_{0}u_{2x}+2iu_{2}\bar{u}_{1}u_{1x}+2iu_{2}\bar{u}_{2}u_{0x}\\&+2iu_{3}\bar{u}_{0}u_{1x}+2iu_{3}\bar{u}_{1}u_{0x}+2iu_{4}\bar{u}_{0}u_{0x},
\end{align*}
$$\vdots $$
Using the  expressions obtained above for equation (\ref{Kom-x2}), we will  illustrate, with two examples, the efectiveness of LADM to solve the nonlinear Kundu-Eckhaus equation.

\section{Numerical Example}

Using Laplace Adomian decomposition method (LADM), we solve this Kundu-Eckhaus equation subject to the initial condition $u(0,x)=f(x)=\beta e^{ix}$. Here $i$ is the imaginary unit and $\beta\in\mathbb{R}$ with $\beta\neq 0$.\\
To use ADM, the equation (\ref{Kom-x2}) is decomposed in the operators (\ref{Oper-1}) and (\ref{Oper-2}).

\noindent Through the  LADM (\ref{eq:ADM1}), we obtain recursively \\
$u_{0}(x,t)=f(x),$\\
$u_{1}(x,t)=\mathcal{L}^{-1}\Big[ \frac{1}{s}\mathcal{L}\{ Ru_{0}+A_{0}\}  \Big],$\\
$u_{2}(x,t)=\mathcal{L}^{-1}\Big[ \frac{1}{s}\mathcal{L}\{ Ru_{1}+A_{1}\}  \Big],$\\
$\vdots$\hspace{0.6in} $\vdots$\\
$u_{n+1}(x,t)=\mathcal{L}^{-1}\Big[ \frac{1}{s}\mathcal{L}\{ Ru_{n}+A_{n}\}  \Big].$\\
Besides
\begin{align*}
A_{0}&=i\beta^{5}e^{ix}, \\
A_{1}&=-\beta^{5}(1-\beta^{4})(3+2e^{2ix})te^{ix},\\
A_{2}&=\beta ^{3}\left( 1-\beta
^{4}\right) \left( 
\begin{array}{c}
4i\beta ^{6}e^{-ix}+11i\beta ^{6}e^{ix}+6i\beta ^{6}e^{3ix}+14\beta
^{4}e^{-ix} \\-\beta ^{4}e^{ix}-14\beta ^{4}e^{3ix}-i\beta ^{2}e^{ix}-7e^{ix}%
\end{array}%
\right)\frac{t^2}{2} \allowbreak ,\\
A_{3}&=
\beta^{3}\left( 1-\beta ^{4}\right) \left( 
\begin{array}{c}
86i\beta^{8}e^{-ix}-143\beta ^{10}e^{-ix}-82\beta ^{10}e^{3ix}\\
-36\beta^{10}e^{5ix}+9i\beta ^{8}e^{ix}-6i\beta ^{8}e^{3ix}\\-96\beta ^{6}e^{-ix} 
+110\beta ^{6}e^{ix}+134\beta ^{6}e^{3ix}\\+36\beta ^{6}e^{5ix}-99i\beta^{4}e^{ix}
-43\beta ^{2}e^{ix}\\-18\beta^{2}e^{3ix}+24ie^{ix}+8i\beta ^{4}e^{-ix} 
\end{array}%
\right)\frac{t^3}{3!} \allowbreak .
\end{align*}
\noindent With the above, we have
\begin{align*} 
u_{0}(x,t) &=\beta e^{ix}, \\
u_{1}(x,t) &= \mathcal{L}^{-1}\Big[ \frac{1}{s}\mathcal{L}\{ iu_{0,xx}+i\beta^{5}e^{ix}\}  \Big]
= \mathcal{L}^{-1}\Big[ \frac{1}{s}\mathcal{L}\{ -i\beta e^{ix}+i\beta^{5}e^{ix}\}  \Big]\\
&= \mathcal{L}^{-1}\Big[ \frac{1}{s^{2}}\Big(i\beta e^{ix}(\beta^{4}-1))\Big) \Big]=-i\beta(1-\beta^{4})te^{ix},\\
u_{2}(x,t)&= \mathcal{L}^{-1}\Big[ \frac{1}{s}\mathcal{L}\{ iu_{1,xx}-\beta^{5}(1-\beta^{4})(3+2e^{2ix})e^{ix}\}  \Big]\\
&= \mathcal{L}^{-1}\Big[ \frac{1}{s}\mathcal{L}\{ \beta(1-\beta^{4})te^{ix}-\beta^{5}(1-\beta^{4})(3+2e^{2ix})e^{ix}\}  \Big]\\
&= \mathcal{L}^{-1}\Big[ \frac{1}{s^{3}}\Big((1-\beta^{4})(\beta e^{ix}-3\beta^{5}e^{ix}-2\beta^{5}e^{3ix})\Big) \Big]\\ &=\big[\beta(1-\beta^{4})(e^{ix}-3\beta^{4}e^{ix}-2\beta^{4}e^{3ix})\big]\frac{t^{2}}{2},
\end{align*}
\begin{align*}
u_{3}(x,t)&= \mathcal{L}^{-1}\Big[ \frac{1}{s}\mathcal{L}\{ iu_{2,xx}+A_2\}  \Big]\\
&=\allowbreak -\beta (1-\beta^{4})\left( 
\begin{array}{c}
4i\beta^{4}e^{ix}+18i\beta^{4}e^{3ix}-ie^{ix}+7\beta^{2}e^{ix}+14\beta^{6}e^{3ix}
+\beta^{6}e^{ix}\\-14\beta^{6}e^{-ix}-6i\beta^{8}e^{3ix}-11i\beta^{8}e^{ix}-4i\beta^{8}e^{-ix}
\end{array}
\right)\frac{t^{3}}{3!},
\end{align*}
\begin{align*}
u_{4}(x,t)&= \mathcal{L}^{-1}\Big[ \frac{1}{s}\mathcal{L}\{ iu_{3,xx}+A_3\}  \Big]\\
&=\allowbreak -\beta (1-\beta^{4})\left( 
\begin{array}{c}
143\beta^{12}e^{-ix}+82\beta^{12}e^{3ix}+36\beta^{12}e^{5ix}-86i\beta^{10}e^{-ix}
\\-9i\beta^{10}e^{ix}+6i\beta^{10}e^{3ix}+100\beta^{8}e^{-ix}
-99\beta^{8}e^{ix}\\-80\beta^{8}e^{3ix}-36\beta^{8}e^{5ix}+100i\beta^{6}e^{ix}-22i\beta^{6}e^{-ix}\\ +126i\beta^{6}e^{3ix}+39\beta^{4}e^{ix}-144\beta^{4}e^{3ix}-17i\beta^{2}e^{ix}
\end{array}
\right)\frac{t^{4}}{4!}.
\end{align*}

\noindent Thus, the solution approximate  of the nonlinear Kundu-Eckhaus equation (\ref{Kom-x2}) is given by:
\begin{equation}\label{Sol-0}
u_{LADM}=u_{0}(x,t)+u_{1}(x,t)+u_{2}(x,t)+u_{3}(x,t)+u_{4}(x,t).
\end{equation}
In the following examples, we will compare (\ref{Sol-0}) with the exact solution of (\ref{Kom-x2}) found in \cite{Arzu},  which is given for $\alpha=1$ as
\begin{equation}\label{exa}
u(x,t)=\pm e^{ix}\cdot\frac{1}{\Big(1+(\frac{1}{u_{0}^{4}}-1)e^{4it}\Big)^{\frac{1}{4}}}
\end{equation}
with the initial condition $u(x,0)=u_{0}e^{ix}$.\\
{\bf Example }
In this  numerical example, we will consider $\beta=\sqrt[16]{2}$. With this value for $\beta$ we obtain
\begin{align*}
u_{LADM}(x,t)&=1.04427e^{ix}+0.19758i te^{ix}-0.19758(-2.3784 e^{3ix}-2.5676 e^{ix})\frac{t^2}{2!}\\&+0.19758 \big(-5.6569 i e^{-ix}+12.920 i e^{3ix}-18.156 e^{-ix}+18.156 e^{3ix}-10.802 i e^{ix}\\&+8.9304 e^{ix}- i e^{ix}\big)\frac{t^3}{3!}
+0.19758 \big(-161.1606 i e^{-ix}+172.6550 i e^{3ix}+381.918 e^{-ix}\\&-146.476 e^{3ix}+9.6329 e^{5ix}+97.2654 i e^{ix}-93.6281 e^{ix}\big)\frac{t^4}{4!}.
\end{align*}
\noindent In Tables \ref{tab:1} and  \ref{tab:2} we  show, for different times,  the values of the exact solution of   (\ref{Kom-x2}) given in \cite{Arzu} and the values given by the approximation previous $u_{LADM}$. The expression for the exact    solution of (\ref{Kom-x2})  is:
\begin{equation}
u_{exact}(x,t)=\frac{e^{ix}}{\Big(1+(2^{-\frac{1}{4}}-1)e^{4it}\Big)^{\frac{1}{4}}}.
\end{equation}
Comparing the values of the exact solution with the values given by LADM,  we  conclude that the approximate solution  is very close to the exact solution  when time values are small. As time becomes greater. \\ 
\noindent The quality of the approximation is also shown in figures 1 and 2 where the real part of the exact  $ u_{exact}$ and the approximate solution $ u_{LADM}$ (imaginary part of $ u_{exact}$ and the approximate solution $ u_{LADM}$ respectively) are depicted in the same figure.   

\begin{figure} [h!]
	\begin {center}
	\includegraphics[width=0.42\textwidth]{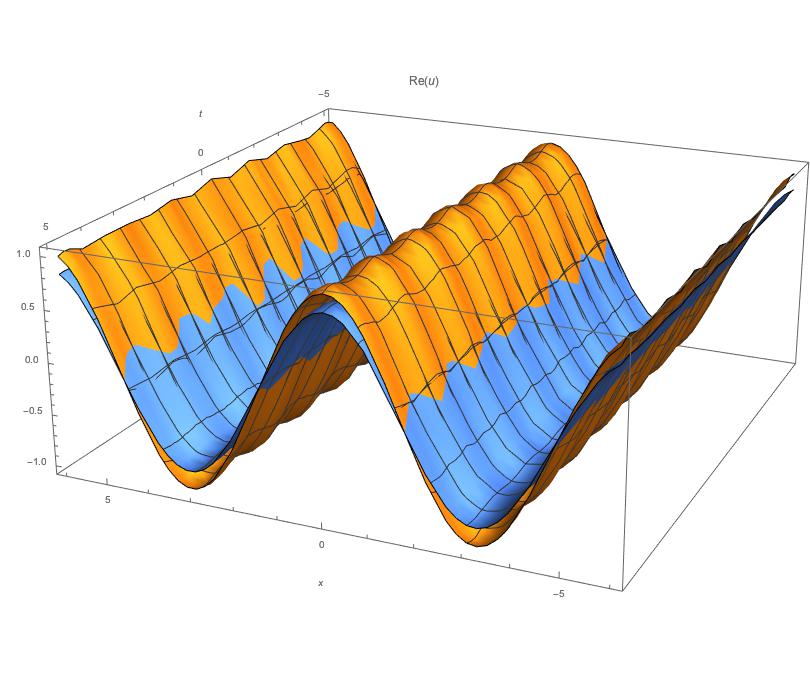}\hspace{0.2in}
	\includegraphics[width=0.44\textwidth]{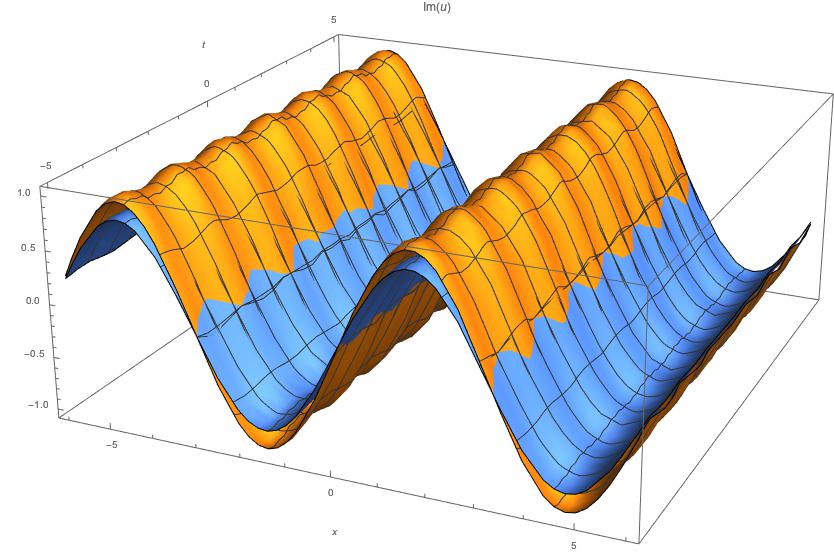}
	\caption{\small Plot of the real part (left)  and imaginary part (right) of the approximate solution $u_{LADM}$ versus the real part of the $u_{exact}$. }  
	\end {center} \label{fi1}
\end{figure}

\begin{figure} [h!]
	\begin {center} \label{fi2}
	\includegraphics[width=0.41\textwidth]{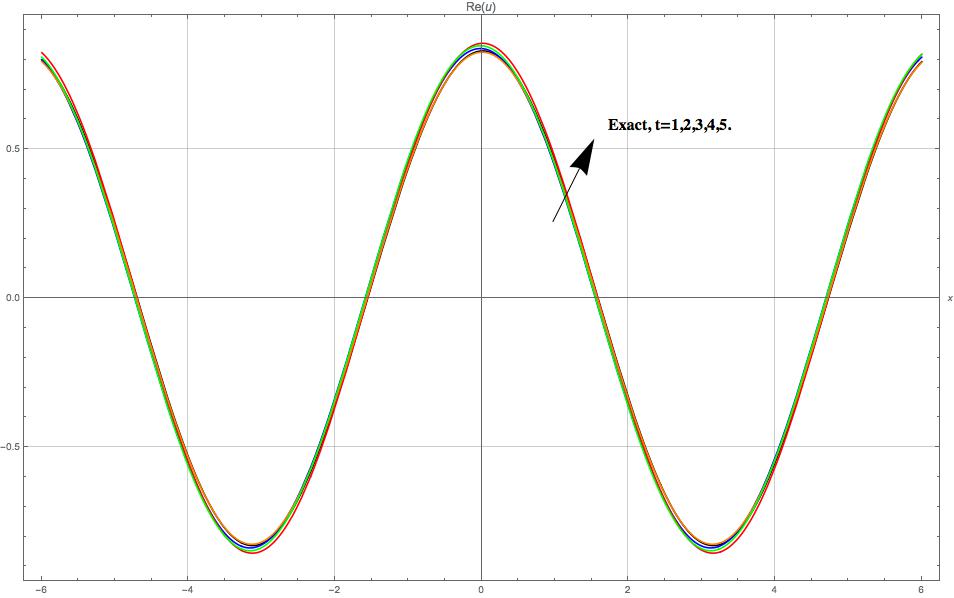}\hspace{0.1in}
	\includegraphics[width=0.41\textwidth]{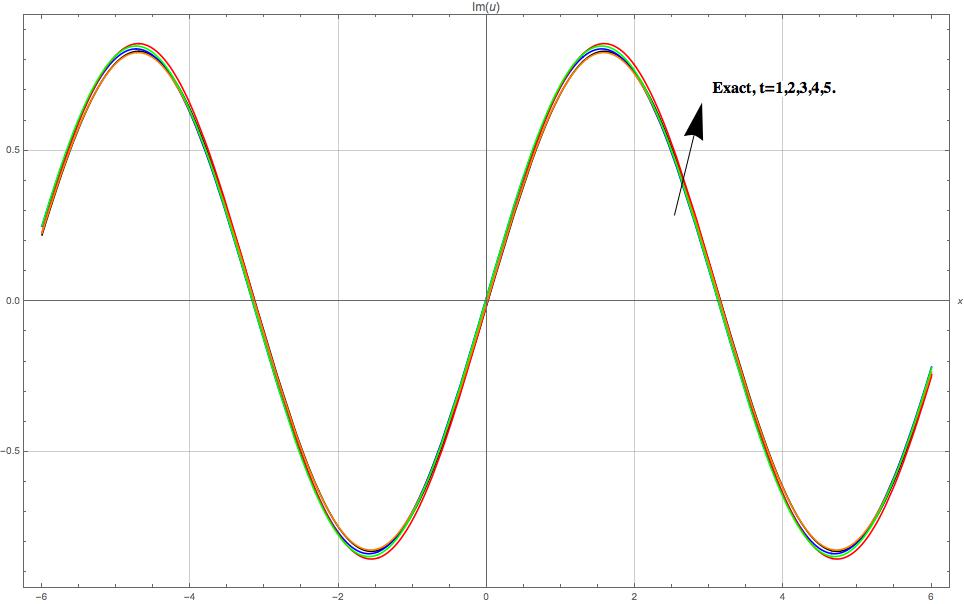}
	\caption{\small Graph of real part (left) and imaginary part (right) of $u_{LADM}$ versus real part  and imaginary part of $u_{ex}$ for $t =1, 2, 3, 4, $ and $ 5 $. }
	\end {center} 
\end{figure}





\noindent From  tables  \ref{tab:1} , \ref{tab:2}, \ref{tab:3} and \ref{tab:4}, we can conclude that the difference between the exact and the obtained LADM approximate solution is very small. This fact tells us about the effectiveness and accuracy of the  LADM method.

\begin{table}[h!]
	\centering
	\tabcolsep=0.2cm
	\begin{tabular}{lllllr}
		\toprule
		\cmidrule(r){1-4}
		$x$  & $ Re(u_{ex}) $ \cite{Arzu}& \hspace{0.1in}$Re(u_{LADM}) $ &  Error\\
		\midrule
		$0.5$ &$0.867245034766$&$   0.867450061289$&$2.05\times10^{-4}$\\
		$1.0$ & $0.548389430252  $&$  0.548531893954$ &$1.42\times 10^{-4} $ \\
		$1.5$ &$0.095268967462 $&$   0.0953139882595 $ &$4.50\times 10^{-5}$   \\
		$2.0$ &$-0.381176661184 $&$  -0.381240105952$ &$6.34\times 10^{-5}$   \\
		$2.5$&$-0.764296949171  $&$  -0.764453326013$&$1.56\times 10^{-4}$ \\
		$3.0$&$-0.960290688213 $&$    -0.960501710624$ &$2.11\times 10^{-4}$ \\
		$3.5$ & $-0.921171775472 $&$  -0.921385777806$&$2.14\times 10^{-4}$ \\
		$4.0$&$-0.656517885107  $&$   -0.656682472129 $&$1.64\times 10^{-4}$  \\
		$4.5$&$-0.231125519605 $&$   -0.231200394673$&$4.78\times 10^{-5}$  \\
		$5.0$&$0.250854433879  $&$  0.250887602795$&$3.31\times 10^{-5}$  \\
		\bottomrule
	\end{tabular}
	\caption{Table for the real parts with $t=1.0$.}
	\label{tab:1}   
\end{table}

\begin{table}[h!]
	\centering
	\tabcolsep=0.2cm
	\begin{tabular}{lllllr}
		\toprule
		\cmidrule(r){1-4}
		$x$  & $ Re(u_{ex}) $ \cite{Arzu}& \hspace{0.1in}$Re(u_{LADM}) $ &  Error\\
		\midrule
		$0.5$ &$0.851260107762$&$  0.851444733853$&$1.84\times10^{-4}$\\
		$1.0$ & $0.503432273480  $&$  0.503521108880$ &$8.88\times 10^{-5} $ \\
		$1.5$ &$0.0323466608364 $&$  0.0323179555396$ &$2.87\times 10^{-5}$   \\
		$2.0$ &$-0.446658542510  $&$  -0.446797760445$ &$1.39\times 10^{-4}$   \\
		$2.5$&$-0.816306156888 $&$   -0.816521802055$&$2.15\times 10^{-4}$ \\
		$3.0$&$-0.986093554388  $&$   -0.986332829329$ &$2.39\times 10^{-4}$ \\
		$3.5$ & $-0.914450858558 $&$   -0.914655180424$&$2.04\times 10^{-4}$ \\
		$4.0$&$-0.618918699965  $&$   -0.619038043636 $&$1.19\times 10^{-4}$  \\
		$4.5$&$-0.171853658076  $&$   -0.171858804059$&$5.14\times 10^{-6}$  \\
		$5.0$&$0.317287152916  $&$   0.317397464537$&$1.10\times 10^{-4}$  \\
		\bottomrule
	\end{tabular}
	\caption{Table for the real parts with $t=2.0$.}
	\label{tab:2}   
\end{table}

\begin{table}[h!]
	\centering
	\tabcolsep=0.2cm
	\begin{tabular}{lllllr}
		\toprule
		\cmidrule(r){1-4}
		$x$  & $ Im(u_{ex}) $ \cite{Arzu}& \hspace{0.1in}$Im(u_{LADM}) $ &  Error\\
		\midrule
		$0.5$ &$0.443634458364$&$   0.443712601886$&$7.81\times10^{-5}$\\
		$1.0$ & $0.805105282409 $&$  0.805272154752$ &$1.66\times 10^{-4} $ \\
		$1.5$ &$0.969458254291$&$   0.969672999286 $ &$2.14\times 10^{-4}$   \\
		$2.0$ &$0.896454034484  $&$  0.896664075067$ &$2.10\times 10^{-4}$   \\
		$2.5$&$0.603966602108 $&$   0.604120513019$&$1.53\times 10^{-4}$ \\
		$3.0$&$0.163607081464 $&$    0.163667179944$ &$6.00\times 10^{-5}$ \\
		$3.5$ & $-0.316809158718$&$  -0.316857586873$&$4.84\times 10^{-5}$ \\
		$4.0$&$-0.719659467741 $&$   -0.719804565630 $&$1.45\times 10^{-4}$  \\
		$4.5$&$-0.946312040059 $&$  -0.946518282658$&$2.06\times 10^{-4}$  \\
		$5.0$&$-0.941274421185 $&$  -0.941491313113$&$2.16\times 10^{-4}$  \\
		\bottomrule
	\end{tabular}
	\caption{Table for the imaginary parts with $t=1.0$.}
	\label{tab:3}   
\end{table}

\begin{table}[h!]
	\centering
	\tabcolsep=0.2cm
	\begin{tabular}{lllllr}
		\toprule
		\cmidrule(r){1-4}
		$x$  & $ Im(u_{ex}) $ \cite{Arzu}& \hspace{0.1in}$Im(u_{LADM}) $ &  Error\\
		\midrule
		$0.5$ &$0.508147216007$&$   0.508299876291$&$1.52\times10^{-4}$\\
		$1.0$ & $0.854056971297  $&$  0.854279457764$ &$2.22\times 10^{-4} $ \\
		$1.5$ &$0.990863793735 $&$   0.991101633938 $ &$2.37\times 10^{-4}$   \\
		$2.0$ &$0.885072601884  $&$  0.885267564246$ &$1.94\times 10^{-4}$   \\
		$2.5$&$0.562584769106 $&$   0.562689120041$&$1.04\times 10^{-4}$ \\
		$3.0$&$0.102356564020 $&$   0.102344754781$ &$1.18\times 10^{-5}$ \\
		$3.5$ & $-0.382932097747 $&$   -0.383057175847$&$1.25\times 10^{-4}$ \\
		$4.0$&$-0.774465626762  $&$   -0.774673350242 $&$2.07\times 10^{-4}$  \\
		$4.5$&$-0.976382959913 $&$    -0.976622470821$&$2.39\times 10^{-4}$  \\
		$5.0$&$-0.939247691931 $&$   -0.939460349642$&$2.12\times 10^{-4}$  \\
		\bottomrule
	\end{tabular}
	\caption{Table for the imaginary parts with $t=2.0$.}
	\label{tab:4}   
\end{table}

\newpage

\section{Conclusions}
\noindent In order to show the accuracy and efficiency of our method,  we have solved two examples, comparing our results with the exact solution of the equation that was obtained in \cite{Arzu}. Our results show that LADM produces highly accurate solutions in complicated nonlinear problems.
\noindent We therefore,  conclude that the Laplace-Adomian decomposition method is a notable  non-sophisticated powerful tool that produces high quality approximate solutions for nonlinear partial differential equations using simple calculations and that attains  converge with only few terms.



\end{document}